\documentclass[traditabstract]{aa} % for the abstract without structuration 
\usepackage{graphicx}
\usepackage{txfonts}
\usepackage{color}
\usepackage{natbib}
\usepackage{tabularx}
\usepackage{multirow}
\usepackage{subfigure}
\def\arcmin{\hbox{$^\prime$} }

\begin{document}

\title{Multi-wavelength environment of the Galactic globular cluster Terzan~5}

\author{A.-C. Clapson\inst{1,2}, W. Domainko\inst{1},
  M. Jamrozy\inst{3}, M. Dyrda\inst{4} and P. Eger\inst{5}}

\institute{Max-Planck-Institut f\"ur Kernphysik, P.O. Box 103980, D 69029 Heidelberg, Germany
\and
EMBL Heidelberg, Meyerhofstrasse 1, D 69117 Heidelberg, Germany
\and 
Obserwatorium Astronomiczne, Uniwersytet Jagiello{\'n}ski, ul. Orla 171,
30-244 Krak{\'o}w, Poland
\and
Instytut Fizyki J\c{a}drowej PAN, ul. Radzikowskiego 152, 31-342 Krak{\'o}w,
Poland
\and
Universit\"at Erlangen-N\"urnberg, Physikalisches Institut, Erwin-Rommel-Str. 1,
D 91058 Erlangen, Germany
}

\titlerunning{Multi-wavelength view of Terzan~5}
\authorrunning{Clapson, Domainko, Jamrozy et al.}

\offprints{\email{clapson@mpi-hd.mpg.de}} 

\date{Received August 10, 2010; accepted May 23, 2011}
 
\abstract
{Terzan~5 is a Galactic globular cluster exhibiting prominent X-ray and
gamma-ray emission. 
Following the discovery of extended X-ray emission in
this object, we explore here archival data at several wavelengths 
for other unexpected emission features in the vicinity of this globular cluster. 
Radio data from the Effelsberg 100 metre telescope 
show several extended structures near Terzan~5,
albeit with large uncertainties in the flux estimates 
and no reliable radio spectral index. 
In particular, a radio source extending from the location of Terzan~5 
to the north-west could result from long-term non-thermal electron 
production by the large population of milli-second pulsars 
in this globular cluster. 
Another prominent radio structure close to Terzan~5 
may be explained by ionised material produced by a field O~star. 
As for the diffuse X-ray emission found in Terzan~5,
its extension appears to be limited to within 2.5 arcmin of the globular cluster
and the available multi-wavelength data is compatible with
an inverse Compton scenario but disfavours a non-thermal Bremsstrahlung origin.
}


\maketitle

\section{Introduction}

\object{Terzan~5} is a Galactic globular cluster (GC) 
located in the inner Galaxy,
at R.A. 17$^h$~48$^m$~04.0$^s$ and Dec -24$^{\circ}$~46'~45" 
(in Galactic coordinates: {\it l} 3.8$^{\circ}$, {\it b} 1.69$^{\circ}$),
at 5.9~kpc from the Sun \citep{ferraro2009}. Terzan~5 hosts 
the largest population of millisecond pulsars (msPSRs) 
detected so far \citep[33, ][]{ransom2008},
and it exhibits the highest central stellar density among Galactic 
GCs \citep{lanzoni2010}.
%, with characteristic radii $r_c = 0^\prime.15$, $r_h = 0^\prime.52$ and $r_t = 4^\prime.6$ 

Recent observational results on Terzan~5 concerning non-thermal signatures
have renewed the interest in this object. 
\citet{eger2010} report diffuse X-ray emission extending beyond 
the half-mass radius of the GC \citep[$r_h=0.52^\prime$, ][]{lanzoni2010}
with a hard spectrum of index close to 1,
likely of non-thermal origin. 
In the high-energy $\gamma$-ray range (100~MeV $<$ E$_{\gamma} <$ 100~GeV) 
Terzan~5 is the brightest GC 
seen with {\it Fermi}-LAT \citep{kong2010,abdo2010}. 
The $\gamma$-ray spectrum of the source is best fitted by a power law 
with an exponential cut-off at a few GeV, as
expected for a population of msPSRs.
\citet{kong2010} also report tentative evidence 
of a second component in the $\gamma$-ray spectrum at energies above
10~GeV, which they interpret as inverse Compton (IC) up-scattering 
of cluster stellar photons by high-energy electrons. 
These observations clearly point to a considerable 
population of non-thermal particles in Terzan~5. 

Several models have been proposed that could explain the observed 
high-energy emission, based for some
on leptons accelerated by the population of 
msPSRs in the GC \citep[e.g. see][]{bednarek2007,venter2008,venter2009}.
These models apply in principle to any GC hosting msPSRs.
The largest predicted fluxes (based on the known msPSRs) are from \object{47~Tucanae}, 
also detected by \emph{Fermi}-LAT \citep{abdo2009} and Terzan~5,
where \citet{abdo2010} estimate a population of 180 msPSRs, 
with a large uncertainty. Since different emission processes
may dominate in the GeV range and at much lower energies,
there is no reason \emph{a priori} for 47~Tucanae to be brighter than
Terzan~5 in all parts of the spectrum.
Another contribution to the high-energy emission
may come from stellar binaries consisting of a normal star and a compact object,
accelerating particles to the required energies 
possibly via non-thermal radio-emitting plasma ejections
\citep[described e.g. in ][]{fender1999}. 
Due to its high core density favourable 
to dynamical interactions between stellar objects, 
Terzan~5 is expected to contain many binary systems 
\citep[e.g.][]{pooley2006,ivanova2008}.
This would also favour merger events,
including white dwarf mergers \citep[see e.g.][]{shara2002} 
and neutron star mergers \citep[discussed in][]{grindlay2006},
expected to result in explosive events, 
type Ia supernovae or short gamma-ray bursts (GRBs) 
depending on the progenitors 
(white dwarfs and neutron stars respectively).
The remnants of these explosions may then be particle-acceleration sites
\citep[see e.g.][]{koyama1995,domainko2005,domainko2008}.

A scenario for extended X-ray emission 
in relation to GCs is proposed in \citet{okada2007}.
They suggest that supersonic motion through the interstellar 
medium (ISM) and bow-shock formation are possible for GCs.
They also find potential signatures
of particles accelerated in GC bow-shocks in radio and X-ray data. 
Typical three-dimensional velocities of GC are around 200~km/s,
but can be as high as 400~km/s for, e.g., NGC~5904 \citep{okada2007}.
Proper motion is unfortunately not available for Terzan~5.
Its radial velocity, a lower limit to the total three-dimensional velocity, 
is about 80~km/s \citep{harris96}.

Previous radio studies of Galactic GCs \citep[see e.g.][]{gopal-krishna1980}
have reported flat-spectrum radio sources,
some of which (\object{M~3} and \object{M~92}) surrounded by arcmin-scale radio features.

Many studies have focused on the GC Terzan~5 itself,
if not on its core. Our aim here, prompted by the discovery of 
X-ray diffuse emission \citep{eger2010}, is 
to collect information from archival data
over spatial scales from the arcmin to the
degree, with particular interest in indications of non-thermal emission
processes.
This paper is organised as follows. First, measurements
from archival data of the region around Terzan~5 are
presented. Radio data are discussed in Section~\ref{section:radio},
 X-ray in Section~\ref{section:xray}, and
infrared in Section~\ref{section:infrared}.
The density of molecular material is investigated in
Section~\ref{section:co} and possible counterparts from 
astronomical catalogues are mentioned in Section~\ref{section:catalogs}.
Then, in Section~\ref{section:processes} the implication of our
findings on the underlying physical processes are discussed.

\begin{figure}[hb]
\centering
\includegraphics[height=12cm]{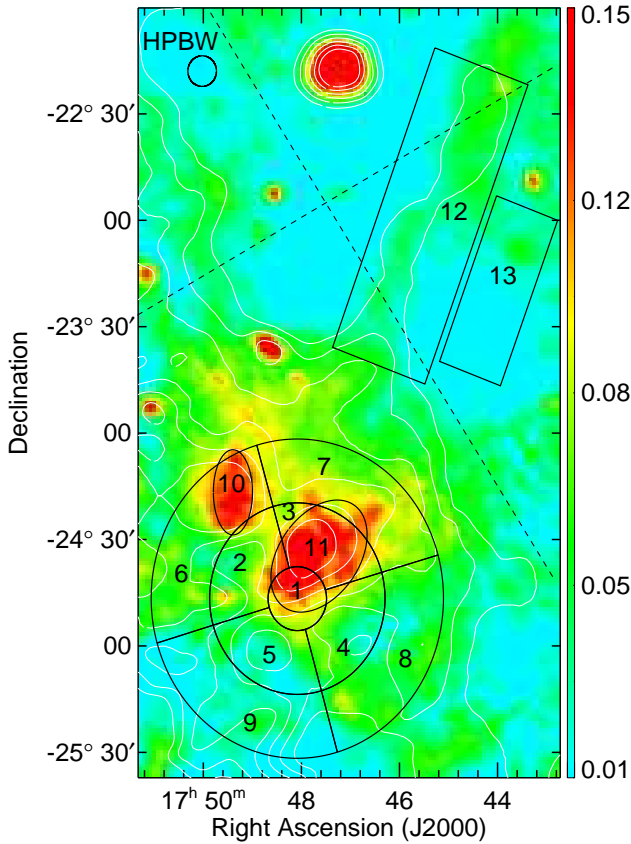}%{Eff11cmallregions.eps}
\caption{Radio map from the Effelsberg Galactic plane surveys 
at 11~cm (in Jy/beam). Contours from the 21~cm map are over-plotted
(for levels 0.2 to 0.7 Jy/beam per 0.1 step),
as well as the regions of Table~\ref{table:regions}.
The region labelled 1 is centred on Terzan~5.
The HPBW for the 11~cm data is also shown.
The dashed lines indicate the radio scan directions at 11~cm.
}
\label{figure:ROIs}
\end{figure}

\section{Radio survey data}\label{section:radio}

\citet{fruchter2000} observed Terzan~5 at wavelengths of 90, 20 and 6~cm
with high angular-resolutions using the VLA in its three most extended 
configurations. Their radio maps display numerous point sources with 
steep spectra, as expected for pulsars. 
Terzan~5 is also detected in the 21~cm NVSS
survey \citep{condon1998} as a 5~mJy single source,
with a weak polarisation feature extending north of the GC.

\subsection{Effelsberg survey data}\label{subsection:effelsbergdata}

For the radio analysis of Terzan~5 we used archival 
21 and 11~cm (respectively 1408 and 2695~MHz) 
total intensity maps from the Survey 
Sampler\footnote {http://www.mpifr-bonn.mpg.de/survey.html}
of the Max Planck Institute for Radio Astronomy. 
These observations were done with the Effelsberg
100-m telescope as part of Galactic plane surveys.
Since we are interested in small-scale structures,
we used the 'source-component' maps, which do not contain
the global steep-spectrum diffuse Galactic emission.
Full details on the data and the processing
can be found in \citet{reich1990a,reich1990b}.
The analysis presented in this paper relies on this preprocessing.
The flux density (S) calibration for the dataset was checked by mapping the
point-like source \object{3C286} and adopting the scale of \citet{baars1977}: 
flux densities of 14.4 (at 21~cm) and 10.4~Jy (at 11~cm).
The angular resolution (half-power beam width, HPBW) of the maps 
are $9^\prime.4$ and $4^\prime.3$, 
their sensitivity (noise rms) 40 and 20~mK~$\mathrm{T}_B$, 
and the ratios $\mathrm{T}_B$[K]$\,$/$\,$S[Jy] are $2.03 \pm 0.04$ and
$2.51 \pm 0.05$, respectively for 21 and 11~cm 
($\mathrm{T}_B$ the brightness temperature). 
The resulting flux maps are shown in Figure~\ref{figure:ROIs} 
and~\ref{figure:objects}.

\begin{table}[hb]
\centering
\begin{tabular}[center]{|*{1}{p{1.cm}}|*{2}{p{1.cm}}|*{2}{p{1.cm}}|*{1}{p{1.cm}}|}
\hline
Region & R$_{min}$ & R$_{max}$ & $\phi_{min}$ & $\phi_{max}$ & Area \\
number & \multicolumn{4}{c|}{[deg]} & [10$^{-6}$~sr] \\
\hline
 1  & 0.0 & 0.15 & 0 & 360 & 22.59 \\
\hline
 2  &  &  & -106 & -16 &  \\
 3  & 0.15 & 0.45 & -16 & 74 & 45.17 \\
 4  &  &  & 74 & 164 & \\
 5  &  &  & 164 & -106 & \\
\hline
 6  &  &  & -106 & -16 & \\
 7  & 0.45 & 0.75 & -16 & 74 & 90.34 \\
 8  &  &  & 74 & 164 & \\
 9 &  &  & 164 & -106 & \\
\hline
10 & 0.1 & 0.2 & 0 & -- & 19.13 \\
11 & 0.2 & 0.3 & 40 & -- & 57.39 \\
\hline
12 & 0.5 & 1.5 & 20 & -- & 228.46 \\
13 & 0.33 & 0.83 & 20 & -- & 84.61 \\
\hline
\end{tabular}
%print,arcminsq/3600./(180./!PI)^2
\caption[Regions]{Description of the regions defined
from the radio maps. Regions 1 is a circle,
regions 2 to 9 are circle wedges, all centred
on the core position of Terzan~5. 
Regions 10 and 11 are ellipses, 
of centres RADec (267.35$^{\circ}$, -24.28$^{\circ}$) 
and (266.91$^{\circ}$, -24.58$^{\circ}$),
major and minor axes R$_{max}$ and R$_{min}$ and inclination $\phi_{min}$.
Region 12 is a rectangular box, centred on RADec 
(266.35$^{\circ}$, -22.98$^{\circ}$), 
of side lengths R$_{min}$ and R$_{max}$ 
and inclination $\phi_{min}$.
The rotation angles are relative to north, westwards. 
\label{table:regions}
}
\end{table}

\subsection{Regions of interest}

The regions used in this work
were defined from visual inspection of the Effelsberg radio data.
Several features can be seen around Terzan~5,
but none exhibits tell-tale morphology
(e.g. shell-like or bow-shock-like structure) 
suggestive of its origin.
Ring segments centred on the GC were defined to cover its environment.
The width of the rings matches the angular resolution of the radio data,
while their orientation aims at isolating main flux levels differences.
Two elliptical regions and a rectangular region 
cover distinct prominent higher flux features, visible in Figure~\ref{figure:ROIs}.
The geometry of all the regions is given in Table~\ref{table:regions}
and shown in Figure~\ref{figure:ROIs}.

The radio flux densities of the regions were extracted 
using the AIPS\footnote{http://www.aips.nrao.edu/} task IRING, or IMEAN for regions~12 and 13. 
The error on the radio flux in a region was derived following
\citet{klein2003}~:
\begin{equation}
\Delta S_i = \sqrt{\left(S_i \times \Delta S_c\right)^2 + (\Delta S_i^n)^2 + (\Delta S_i^z)^2}
\end{equation}
\noindent for $S_i$ the flux measurement at frequency $\nu_i$ 
and $\Delta S_i$ the related error. We used a calibration error $\Delta S_c$ of 3\% 
and a noise rms $\Delta S_i^n$ derived from the rms noise of the map
(respectively 0.08 and 0.05 Jy/beam at 21 and 11~cm) 
and multiplied the results by the integration area.
The estimation of a `diffuse-component' performed 
by \citet{reich1990a, reich1990b} may locally produce large inaccuracies. 
The 'diffuse-component' was subtracted from the original data to obtain 
the 'source-component' map used here. 
We consider a zero-level error $\Delta S_i^z$ at 3\% of the mean flux value 
in the region of the `diffuse-component' map. 
$\Delta S_i^z$ also depends on the area 
(in units of solid angle, used to determine the zero-level). 
$\Delta S_i^z$ is the predominant contribution to the flux errors.

The flux density values are summarised in Table~\ref{table:values}.
Regions~4, 5 and 8 might indicate local background levels at 11 and 21~cm, 
with a possible deviation for region~9 (below at 11~cm, above at 21~cm).
While two bright regions (10 an 11) stand out at both wavelengths,
it is difficult to make out other features, in particular at 21~cm,
where the errors are larger and the resolution worse.
Regions~1, 2, 3, 6 and 7 overlap somewhat with the bright features
and present as expected intermediate flux values. 

Region~12 is visible
only because of the very low flux densities in its surroundings.
Since it is located at higher galactic latitudes the background level of the
radio emission is considerably lower than close to Terzan~5.
To quantify the robustness of the structure we estimated the background
level of the flux density in region~13, adjacent to region~12 at the
same galactic latitude (see Table~\ref{table:values}). At least at 21~cm,
region~12 stands significantly above the local background level.

\subsection{Spectral index estimation}

We extracted the mean spectral index $\alpha$ for each region from the two flux
measurements in each region. The uncertainty on $\alpha$,
estimated following equation~3 of \citet{odea2009} 
%\begin{equation}
%\Delta \alpha = \frac{1}{ln (\nu_1 / \nu_2)} \sqrt{\left(\frac{\Delta S_1}{S_1}%\right)^2 + \left(\frac{\Delta S_2}{S_2}\right)^2}
%\end{equation}
%\noindent 
is very large in all regions. We therefore do not present the radio spectral
information in this paper. For the discussion we use a spectral index of -0.8.
Detailed re-analysis of the archival data 
or additional data spanning a wider frequency range may 
reduce this limitation.

\section{X-ray data}
\label{section:xray}

\subsection{\emph{ROSAT}}

The first detection of extended diffuse emission from the GC 
Terzan~5 was reported by \citet{eger2010}
who analysed the data of an archival \emph{Chandra} 
Advanced CCD Imaging Spectrometre (ACIS) observation.
Following this discovery we investigated archival \emph{ROSAT} data, to
search for diffuse emission on a much larger spatial scale
than possible with the limited field of view (FoV) of \emph{Chandra}.
Terzan~5 was observed by \emph{ROSAT} several times \citep[see e.g.][]{johnston1995}. 
We focused on a pointing observation centred on Terzan~5
taken in 1991 (ID 300060)
by the High Resolution Imager (HRI) instrument
with a total live time of 22.5~ks. 
The advantage of the \emph{ROSAT}-HRI observation
compared to \emph{Chandra} is its large square FoV
with a side of 38$^\prime$. 
Data analysis relied on the standard packages FTOOLS 5.9 from HEASOFT 
version 6.8 \citep{blackburn1995} 
and CIAO\footnote{http://cxc.harvard.edu/ciao/}.

To search for extended diffuse emission,
point-like sources must first be excluded. 
There are 50 Chandra point-sources inside $r_h$, 
listed in Table~1 of \citet{heinke2006}. 
To account for the difference between the 
point spread function (PSF) of \emph{ROSAT}-HRI 
\citep[$\sim$ 5$^{\prime\prime}$,][]{boese2000}
and \emph{Chandra}-ACIS \citep[$\sim$ 1$^{\prime\prime}$, ][]{karovska2001}, 
we defined a large exclusion region at the centre of the GC, 
with a radius equal to $r_h$.
Outside $r_h$, \citet{heinke2006} found 78 \emph{Chandra} sources
(listed in their Table~2). 
We excluded them taking into account the HRI PSF. 
All the excluded regions were removed from the dataset 
and refilled with the CIAO DMFILTH tool, assuming Poisson statistics. 

\begin{figure}[hb]
\centering
\includegraphics[height=7cm]{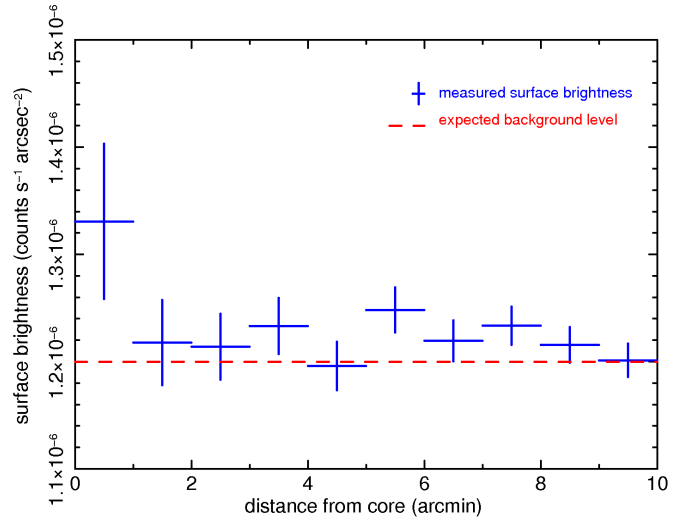}%{HRI_surface_flux.eps}
\caption{\emph{ROSAT}-HRI surface brightness 
versus distance from the core position of Terzan~5, with a 1~$\sigma$ error bar.
The expected background level is represented by the dashed horizontal line. 
The extraction regions are adjacent rings of equal width, see text for details.}
\label{figure:hri_surface_flux}
\end{figure}

Since the spectral resolution of the HRI camera is very limited,
we only extracted the total count-rate in the full HRI energy band
(0.1~--~2.4~keV). 
The HRI background is discussed in the \emph{ROSAT}-HRI
calibration report \citep{david1997}.
It combines different components: the internal background, 
the externally-induced background from cosmic rays 
and the X-ray background (Galactic and extra-galactic).
For our analysis we adopted the given standard values 
for the first two (pointing-independent) components. 
Terzan~5 is located near the Galactic plane, 
therefore the X-ray background, in particular 
the diffuse Galactic emission, is larger than average 
over the sky. Based on the results from \citet{eger2010}, 
we estimated this component as
3$\times10^{-7}$~counts$\,$s$^{-1}\,$arcsec$^{-2}$.
Compared to \citet{david1997}, 
this is slightly above the standard value but still 
within the given range. 
The total expected HRI background level for this observation amounts
then to $1.2\times10^{-6}$~counts$\,$s$^{-1}\,$arcsec$^{-2}$ 
in the 0.1~--~2.4~keV energy band. 
  
Diffuse emission connected to Terzan~5 was searched for using 
concentric annular extraction regions 1\arcmin thick
centred on the GC, of internal radii 0 to 10$^\prime$. 
We integrated the counts within each ring from the refilled HRI image
and compared these to the afore-mentioned background component level. 
Figure~\ref{figure:hri_surface_flux} shows the surface brightness 
for the rings. 
All the extracted values agree within 1~$\sigma$ and
are compatible with the expected background.
The innermost region, somewhat brighter than the others,
is still within less than 2$\sigma$ deviation from the expected background.
The diffuse emission detected with
\emph{Chandra} \citep{eger2010} as well as un-removed point sources are
likely to contribute to the flux measured in this innermost ring.
In addition there might still be some contamination from the wings 
of the PSF of removed point-sources or a contribution from 
a transient source, a common phenomenon 
in dense GCs like Terzan~5 \citep[see e.g.][]{heinke2002}. 

The archival HRI data confirm the previous results of \citet{eger2010}
and show no indication of emission above the Galactic diffuse component 
beyond $2^\prime.5$ away from Terzan~5. 
The extended X-ray emission connected to Terzan~5
seems therefore to be very localised, 
a result not accessible with the small-FoV \emph{Chandra} data
analysed in \citet{eger2010}.

\subsection{\emph{INTEGRAL}}

The non-thermal X-ray emission detected by \citet{eger2010} might be
detectable in the hard X-ray regime by \emph{INTEGRAL} 
if it extends to this energy band. 
For this source, of un-absorbed flux
5.5$\times10^{-13}$~erg$\,$cm$^{-2}\,$s$^{-1}$ 
and spectral index of about 0.9 in the 1~--~7~keV band \citep{eger2010}, 
a flux of 4.9$\times10^{-12}$~erg$\,$cm$^{-2}$s$\,^{-1}$ in the 17~--~60~keV
band is expected. This might be within the reach of \emph{INTEGRAL}, 
which can detect sources down to
3.7$\times10^{-12}$~erg$\,$cm$^{-2}\,$s$^{-1}$ in the same band 
\citep{krivonos2010} near the Galactic centre 
(and also at the position of Terzan~5).

\section{Other wavelengths}

Infrared and $^{12}$CO survey data were also studied,
to gather information about the environment from dust
and molecular gas content.
A gradient in infrared emission in the direction perpendicular 
to the Galactic plane was found.  
For the whole region a very low average molecular gas column density
was measured.
No significant local features were found that could hint of peculiar 
conditions in the ISM close to the GC. 

\subsection{Far infrared data}\label{section:infrared}

The Infrared Astronomical Satellite (\emph{IRAS}) archival data,
here from the IRIS data reduction chain \citep{miville2005},
provides full-sky coverage at 4 wavelengths (12, 25, 60 and 100~$\mu$m).
The angular resolution (4.2\arcmin at 100~$\mu$m) 
is sufficient to distinguish differences on the scale of the regions
considered here.
Bright point-like sources from the \emph{IRAS} 
catalogue \citep{beichman1988} were removed
when estimating the average flux in the regions.
After comparison to \emph{Spitzer} 
maps\footnote{http://irsa.ipac.caltech.edu/applications/Cutouts/}, 
the limiting flux for source exclusion
was set in the 12~$\mu$m band, where the contribution from stars is largest, 
to exclude at least the two \emph{IRAS} point-like sources\footnote{Catalogue accessed 
from Vizier \citep{ochsenbein2000}.} closest to Terzan~5,
corresponding to the GC itself and a nearby bright star.

A flux density gradient is observed across the region,
with a factor 2 decrease when moving away from the Galactic plane,
with the lowest fluxes measured for region~12, the furthest away 
from the Galactic plane.
Visual inspection of \emph{IRAS} and \emph{Spitzer} maps
did not reveal any obvious extended structure in this region
apart from Terzan~5 itself, where star crowding would mask
any truly extended emission component.

\subsection{Molecular gas and $^{12}$CO data}\label{section:co}

The largest-coverage molecular-gas-tracer map
of the Galactic plane was compiled by \citet{dame2001}, hereafter DA01, 
from the data of several surveys
of the CO rotational transition 1~--~0 emission line.
We extracted the temperature values against velocity from their publicly 
available\footnote{http://www.cfa.harvard.edu/mmw/obtainingData.html}
data cube.
The values were integrated over 
the relevant velocity range, over the spatial extension of the region
and finally converted into an hydrogen column density
$\mathrm{n}_{\mathrm{H}_2}$ (cm$^{-2}$) using
the conversion factor 1.8$\times10^{20}$~cm$^{-2}\,$K$^{-1}\,$km$^{-1}\,$s
from DA01. 

\begin{figure}
\centering
\includegraphics[width=9cm]{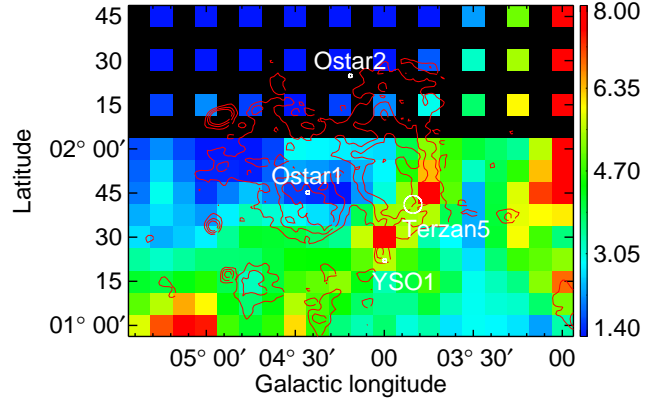}%{CO12_3to10kms_objects.eps}
\caption{Map of $\mathrm{n}_{\mathrm{H}_2}$
derived from the $^{12}$CO emission in the 3 to 10~km/s velocity range
(roughly equivalent to a distance range of 1 to 3~kpc) 
overlaid with contours from the 11~cm data 
(for levels 0.07, 0.095 and 0.12~Jy/beam)
and selected objects (see Section~\ref{section:catalogs}).
The Terzan~5 circle indicates the half-mass radius of the GC.  
Colour scale normalised by 10$^{20}$~cm$^{-2}$. 
The map was blanked where no value was available.
}
\label{figure:COmap}
\end{figure}

Using the close-case distance estimates 
from a Galactic rotation model 
\citep[here following][]{clemens1985,nakanishi2003}, 
we selected the boundary velocities
to bracket the distance to Terzan~5 (5.9~kpc): 
20~km/s ($\approx$5~kpc) and 120~km/s ($\approx$7.3~kpc). 
The complete range provided by DA01
was also considered (from minus to plus 170~km/s).
The estimated $\mathrm{n}_{\mathrm{H}_2}$ are very low,
as illustrated in Figure~\ref{figure:COmap}, 
compatible with zero in some regions,
notably region~12 in all velocity ranges.
The maximum values are measured in regions~6 and 9,
with respectively 2.6 and 4.5$\times10^{21}$~cm$^{-2}$
in the full velocity range and 
3.4 and 3.3$\times10^{20}$~cm$^{-2}$ in the velocity range
compatible with the distance to Terzan~5.
These values are comparable to other regions at similar Galactic latitude,
where the latitude gradient of $\mathrm{n}_{\mathrm{H}_2}$ was shown to be large (DA01).
The neighbouring giant molecular cloud Bania~Clump~2
mentioned in DA01 does not extend to this region.
Most of the measured column density is associated to velocity range $v < $20~km/s
(distances below 5~kpc). We interpret it as the cumulative emission
from the diffuse molecular gas near the Galactic plane along the line of sight.
Marginally significant structures found 
from visual inspection of the data cube
are discussed in Section~\ref{section:catalogs}.
The slightly larger values of $\mathrm{n}_{\mathrm{H}_2}$ for regions~6 and 9 
correspond to the shorter distance of these regions to the Galactic plane.
Region~12 is the furthest away from the Galactic plane, away
from distinct high-latitude features, so the null result (to the sensitivity
of the available data) is not surprising.

\begin{table*}
\centering
\begin{tabular}[center]{|*{1}{p{1.9cm}}|*{13}{p{0.75cm}}|}
\hline
 & \multicolumn{13}{c|}{region} \\
 & 1 & 2 & 3 & 4 & 5 & 6 & 7 & 8 & 9 & 10 & 11 & 12 & 13 \\ 
\hline
Wavelength & \multicolumn{13}{c|}{Average radio flux density (error)} \\
{[cm]} & \multicolumn{13}{c|}{[kJy/sr]} \\
\hline
% Uses 11cmconv values
11 & 
62.60 & 39.56 & 64.44 & 28.16 & 
18.31 & 41.39 & 45.06 & 25.56 & 8.41 & 
61.58 & 67.21 & 12.89 & 7.00 \\
 &
(9.39) & (6.87) & 
(6.43) & (6.35) & (6.94) & 
(5.10) & (4.42) & (4.35) & 
(5.21) & (10.20)& (5.85) &
(2.04) & (6.94) \\ 
\hline
21 & 
60.88 & 63.76 & 74.05 & 33.94 & 
40.69 & 81.75 & 55.71 & 35.54 & 50.63 & 
86.67 & 73.53 & 18.54 & 1.29 \\
 &
(17.74) & (12.73) & 
(11.84) & (11.93) & (12.91) & 
(9.47)  & (8.00)  & (8.23)  & 
(9.66)  & (18.97) & (10.67) &
(4.07) & (4.30) \\ 
\hline
\end{tabular}
\caption[values]{Measurements extracted for the regions from 
the Effelsberg radio maps \citep{reich1990a,reich1990b}.
Regions~4, 5, 8 and 13 might indicate local background 
levels at 11 and 21~cm.
\label{table:values}
}
\end{table*}

\section{Catalogue search}
\label{section:catalogs}

\begin{figure}
\centering
\includegraphics[height=12cm]{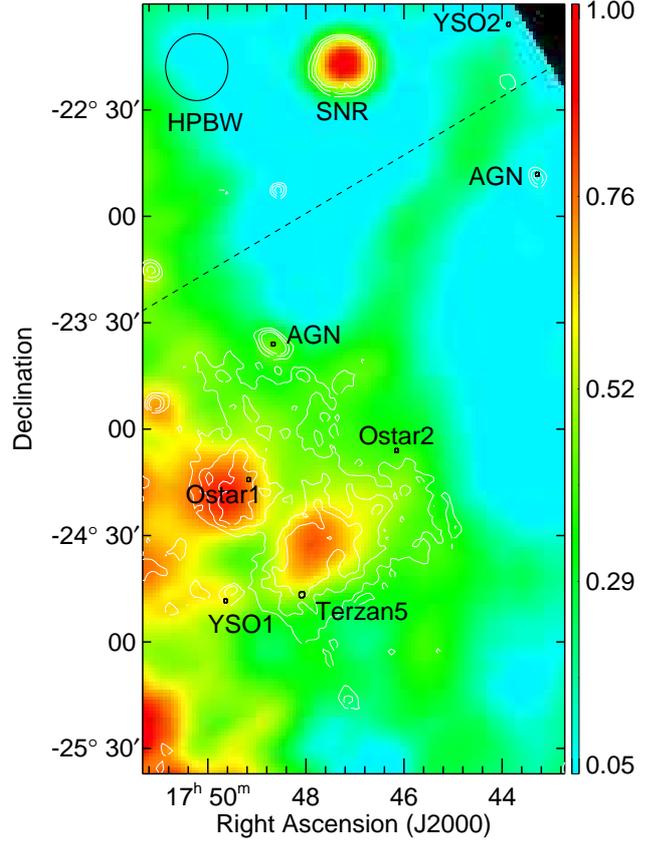}%{Eff21cmcontour11cm.eps}
\caption{\label{figure:objects}
Radio map from the Effelsberg Galactic plane surveys 
at 21~cm (in Jy/beam). Contours from the 11~cm map
(for levels 0.07, 0.095 and 0.12~Jy/beam), 
selected catalog objects as well as the HPBW 
for the 21~cm data are over-plotted.
The Terzan~5 circle indicates the half-mass radius of the GC.
The dashed line indicate the radio scan direction at 21~cm.
}
\end{figure}

Previously identified objects potentially relevant to our study
were searched with the help of the CDS online tools\footnote{http://cdsweb.u-strasbg.fr/}.
There appears to be no particularly energetic object in the region
that could straightforwardly generate the entire observed radio emission.
A few objects are worth mentioning (shown in Figure~\ref{figure:objects}). 
The large radio source east of region~12 is \object{SNR WR143}.
Two active galactic nuclei have been identified, one on each side of region~12,
both as bright radio point sources.
Two O~type stars are found, one near the centre of region~10, the other 
north of Terzan~5. Both stars seem to be 
at a distance below 2~kpc \citep{reed2005}.
Figure~\ref{figure:COmap} shows $\mathrm{n}_{\mathrm{H}_2}$ 
for a compatible velocity range, 
with a local minimum coincident with one of the stars, labelled 'Ostar1' 
in the figure and catalogued as \object{2MASS~J17491016-2414211}.
This may hint of a previous interaction between the star and molecular gas
leading to a cavity. The data is too sparse near the other star 
to make any statement. Based on the \emph{IRAS} source 
catalogue
%\footnote{\emph{IRAS} catalogue of Point Sources, Version 2.0 (IPAC 1986)} 
\citep{beichman1988} and young stellar object (YSO) selection criteria 
from \emph{IRAS} colours \citep{junkes1992}, two candidate YSOs were identified, 
one of them (labelled as 'YSO1') close to a radio hot spot 
and possibly a $^{12}$CO one.
All the other bright radio hot spots were previously catalogued \citep{reich1990a}
but we found no additional published information on them.

\section{Constraints on non-thermal emission processes}
\label{section:processes}

Based on the collected measurements,
this Section focuses on testing non-thermal radiation processes 
that may generate the emission around Terzan~5. 
The energetics of the
detected radio structures are constrained
in Section~\ref{subsection:SRenergetics}. 
The particular properties of regions~10, 11 and 12 are
discussed in Sections~\ref{subsection:ROI10} to
\ref{subsection:ROI12}.
Potential associations between Terzan~5 and region~11 or 12 are discussed.
Section~\ref{subsection:diffuseX} 
examines the origin of 
the diffuse X-ray emission found by \citet{eger2010}.  

While a physical connection between the structures 
seen around the direction of Terzan~5 
(Figures~\ref{figure:ROIs} and \ref{figure:objects})
and the GC itself cannot be affirmed, 
it is assumed hereafter
that they are all located at the same distance of 5.9~kpc.
For the interpretation of the radio data, it is further assumed 
that synchrotron radiation (SR) emission is the dominant scenario.
If the electron energy distribution 
follows a power law of index $p$, $N(E) dE = \kappa E^{-p} dE$,
the radio spectral index $\alpha$ of the SR emission is $\alpha = (p-1)/2$. 
Typical values \citep{longair1992} for $\alpha$ are: 
for centre-filled supernova remnants (SNR) $\alpha \approx $~-0.3~--~0.0, 
for SNR shells $\alpha \lesssim -0.5$ and 
for the Galactic radio GHz range emission $\alpha \approx $~-0.9~--~-0.8. 
In the absence of radio index measurements, we assume for all regions $\alpha = -0.8$.
Along with this, a Bremsstrahlung scenario is investigated for region~10.

\subsection{Synchrotron radiation energetics}\label{subsection:SRenergetics}

To estimate the energy in the radio structures assuming SR emission, 
the minimum energy condition (MEC) is used, 
as reviewed in \citet{miley1980}.
The MEC density of a radio source can be written as
$u_{me} = (7/3) (B^2_{me}/8\pi)$~erg$\,$cm$^{-3}$
and corresponds almost (but not exactly) 
to the equi-partition of energy between non-thermal particles and 
the magnetic field. For the frequency interval 
$\nu = 0.01$~---~100~GHz and assuming a spectral index $\alpha = -0.8$,
the corresponding magnetic field, in $\mu$G (for frequencies in GHz), is given by:
\begin{equation}
B_{me} = 140 \quad \left(\frac{\nu_0}{1 \mathrm{GHz}}\right)^{0.22} \quad \left[ \frac{S_0}{(1 \mathrm{Jy})} \frac{(1 \mathrm{arsec}^2 \, 1 \mathrm{kpc})}{\theta_x \, \theta_y \, s}\right]^{2/7}
\label{equation:Bminenergy}
\end{equation}
\noindent where $\theta_x$ and $\theta_y$ are the extent of the radio source 
on the plane of the sky in orthogonal directions, 
$s$ is the path length through the source along the line of sight
and $\nu_0$ is the frequency where the flux $S_0$ is measured.

The MEC is assessed from the measurements
of Table~\ref{table:values}, for $\nu_0 = 2.7$~GHz. 
It is assumed that region~1 is spherically symmetric 
with a radius of 15~pc and that regions~3 to 10 extend by the same 30~pc 
along the line of sight ($s =$ 30~pc). 
The regions~10 to 12 are assumed to be structures where $s$ equals 
the minimum extension of the source in the plane of the sky,
estimated here for a distance of 5.9~kpc. 
Since the strength of $B_{me}$ scales only with $(S_0 / s)^{2/7}$,
the uncertainties on the actual path lengths and the flux measurements are not critical
to derive orders of magnitude. 
Under these assumptions, energetics for all the regions
fall between a few 10$^{48}$~ergs and more than 10$^{49}$~ergs. 
The values for $u_{me}$, $B_{me}$ as well as the total energy $E_{me}$
(in particles and in the magnetic field) in each region 
are given in Table~\ref{table:energetics}. 
These values serve as references in the following region-by-region discussions.

\begin{table*}
\centering
\begin{tabular}[center]{|*{1}{p{2.7cm}}|*{12}{p{0.7cm}}|}
\hline
 & \multicolumn{12}{c|}{region} \\
 & 1 & 2 & 3 & 4 & 5 & 6 & 7 & 8 & 9 & 10 & 11 & 12 \\ 
\hline
$B_{me}$ [$\mu$G] & 10.3 & 8.9 & 10.3 & 8.1 & 7.2 & 9.1 & 9.3 & 7.9 &
5.7 & 11.4 & 9.5 & 5.6 \\
$u_{me}$ [$10^{-12}$~erg$\,$cm$^{-3}$] & 9.8 & 7.4 & 9.8 & 6.1 & 4.8 &
7.7 & 8.0 & 5.8 & 3.0 & 12 & 8.4 & 2.9 \\
$E_{me}$ [$10^{48}$~erg] & 2.1 & 8.1 & 11 & 6.7 & 5.3 & 19 & 20 &
15 & 7.5 & 2.8 & 12 & 29 \\
\hline
\end{tabular}
\caption{Magnetic field strength, energy density and energetics for
the regions computed according to the MEC. (see text for details).
\label{table:energetics}
}
\end{table*}

\subsection{The O~star and region~10}\label{subsection:ROI10}

Region 10 is located north-east of Terzan~5.
It does not overlap with the GC
but contains an O~star at a distance shorter than 2~kpc (see Section~\ref{section:catalogs}). 
Even though an association with the GC cannot be excluded, 
we now assume that region~10 is related to the O~star,
which could serve as source of ionising radiation 
for a Bremsstrahlung scenario.
There are indications for a hole in the $^{12}$CO distribution 
at distances between 1 and 3~kpc consistent with 
the location and potentially the distance of the O~star. 
In the Bremsstrahlung picture the O~star would ionise 
the embedding molecular material causing this hole in the 
$^{12}$CO distribution. The ionised material would 
then emit Bremsstrahlung 
\citep[e.g.][]{vasquez2010}. In such a scenario, n$_e$ would 
be comparable to the molecular gas density, of the order of 1~--~10~cm$^{-3}$.
A flux density from region~10
of 14~$(T/10^4\, \mathrm{K})(n_e/10\, \mathrm{cm}^{-3})$~Jy 
would be expected, in broad agreement with the observed value. 
The expected Bremsstrahlung spectrum would be flat in the radio range
($\alpha \approx \, -0.1$). As visible in
Figure~\ref{figure:objects}, the features at 11 and 21~cm 
corresponding to region~10 do not match exactly.
The possibility remains therefore that the set of properties
used here for this region do not accurately describe it.
Observation at higher spatial resolution may settle this issue.

\subsection{Accumulation of relativistic electrons in region~11}
\label{subsection:ROI11}

The enhanced emission in region~11 covers the location of
Terzan~5 and extends from there to the north-west of the GC,
away from the Galactic Plane.

\subsubsection{Evidence from the radio maps}

The radio emission in region~11 represents 
in the MEC a total energy of $9.2 \times 10^{48}$~ergs. 
For SR emission peaking at a few GHz in the MEC magnetic field 
of about 10~$\mu$G (see Table~\ref{table:energetics}), 
electrons with an energy of about 10 GeV would be 
required \citep[see e.g.][]{longair1992}. Following \citet{aharonian2004},
the cooling time of these electrons in such a magnetic field 
would be about 10$^7$~years. 
This scenario could be tested by the radio index, as
radio spectral indices of -0.3~--~0.0 are typical of pulsar wind nebulae 
\citep[PWNs, see][]{gaensler2006}. Sampling the radio index within region~11
could then be used to estimate the electron production time scale,
expected to be shorter than the cooling time if no spectral steepening
was found.

\subsubsection{Possible association with Terzan~5}

Given the geometry of the radio feature covered by region~11, 
the large population of msPSRs in the GC 
could be the source of the radio-emitting electrons. 
Assuming that Terzan~5 contains 180~msPSRs \citep{abdo2010},
each injecting energy on average at a moderate rate of 10$^{33}$~ergs/s 
in electrons into the surroundings, about 2$\times 10^6$~years 
would be needed to build up the total electron population. 
This time scale is shorter than the cooling time estimated above,
so this scenario seems viable. 

The radio structure covered by region~11 extends from Terzan~5
in a direction roughly perpendicular to the Galactic plane. 
Based on the energy density given in Table \ref{table:energetics} 
and assuming for the ISM a density of 0.1~cm$^{-3}$ 
and a temperature of 10$^5$~K \citep[see e.g.][]{savage1981}, 
this radio structure would seem to be mildly over-pressured 
(by a factor of 3) with respect to the surrounding gas.
It could therefore be expected to expand,
preferentially in the direction of the Galactic
density gradient, as observed. 
In this scenario, the expansion of the bubble is driven by the 
difference between the internal and external pressures 
($p_{int}$ and $p_{ext}$) over the surface $S$ where they exert,
with $F_{p} \sim (p_{int} - p_{ext}) \times S$.
It is limited by the ram pressure of the surrounding 
medium, $F_{ram} \sim C \frac{1}{2} S v^2 \rho_{ext}$
according to \citet{churazov2001},
with $C$ the drag coefficient, $v$ the expansion velocity and $\rho_{ext}$ 
the density outside the bubble. The maximum terminal expansion
velocity is then given by:
\begin{equation}
v \sim \sqrt{\frac{2}{C} \frac{p_{int} - p_{ext}}{\rho_{ext}}}
\end{equation}
\noindent For $C=0.75$ \citep{churazov2001}, 
$v \sim 80\,(n/0.1\mathrm{cm}^{-3})$~km/s. At this constant velocity,
of the order of 10$^6$~years are needed to expand by 100~pc. 
These numbers are in rough agreement with the production timescale 
of the electron population in region~11.

Another mechanism operating at the location of Terzan~5
could potentially displace radio-emitting plasma produced in the GC 
in the direction perpendicular to the Galactic plane. 
Terzan~5 is placed in the onset region of the Galactic wind 
of our Galaxy, where the bulk motion
flow of the wind material is expected to exceed 100~km/s \citep{everett}. 
This Galactic wind could push away material originating in Terzan~5 
even faster than the pressure
imbalance discussed in the previous paragraph.

Detection of radio polarisation in this region,
challenging at the distance of Terzan~5 
\citep[see e.g. the discussion on polarisation horizon in][]{landecker2010}, 
would support a plasma plume scenario.
These dynamical models could be further tested against the proper 
motion of the GC.
For instance, in $2 \times 10^6$~years and for a velocity in the plane 
of the sky of 50~km/s, a 100~pc long structure could be generated
in the direction of motion. Conversely,
too high a velocity or an incompatible direction of proper motion would
exclude the proposed association between the radio feature
in region~11 and Terzan~5.

\subsection{The large-scale ridge in region~12}\label{subsection:ROI12}

The ridge-like radio structure seen in region~12 
stands out for its length and apparent alignment with Terzan~5.
It is more visible in the 21~cm map (see Figure~\ref{figure:objects}).
Structures extending over similar linear
scales and involving comparable energetics
have been proposed \citep[][]{fukui2009,yamamoto2008}
to result from highly collimated outbursts from either
micro-quasars or stellar explosions.
Such objects may be hosted by GCs like Terzan~5 but
no equivalent central object or active region could be identified 
in relation to region~12. Higher sensitivity observations and dedicated investigations
appear to be necessary to establish a full emission scenario for this region.

\subsection{Revisiting the diffuse X-ray emission around Terzan~5}
\label{subsection:diffuseX}

\subsubsection{Non-thermal Bremsstrahlung}

The molecular gas and infrared data can be used to test a non-thermal 
Bremsstrahlung origin for the extended diffuse X-ray emission in Terzan~5. 
In this scenario, electrons are deflected by target nuclei from
the ambient medium. Therefore the intensity of the non-thermal X-ray 
emission should be correlated with the density of the local ISM. 
The absence of observational support from the $^{12}$CO or infrared data 
for regions with high density of target material goes toward excluding 
a non-thermal Bremsstrahlung scenario for the diffuse X-ray emission 
in Terzan~5. More detailed investigation of the infrared data
as well as higher resolution molecular gas tracer data may
change this conclusion.

\subsubsection{Inverse Compton scenario}

One possible origin for the diffuse
X-ray emission discovered by \citet{eger2010}
is X-ray IC radiation, usually accompanied 
by radio SR emission. 
The data gathered in this work can constrain this scenario. 

The predicted radio SR emission is described by the frequency 
$\nu_{syn} = 120\,(\gamma/10^4)^2\,(B/1 \mu\mathrm{G})\,$sin$\phi$~MHz,
with $\gamma$ the Lorentz factor of the electrons, 
$B$ the magnetic field and sin$\phi$ the pitch angle 
of the electrons with respect to the magnetic field. 
For $\gamma \approx 30$, typical of IC radiation in the keV 
range \citep[see][]{krockenberger1995}, and neglecting sin$\phi$, 
one gets $\nu_{syn}=1.2\,(B/1 \mu\mathrm{G})$~kHz. 
The SR flux relates to the IC flux 
by $F_\mathrm{syn}/F_\mathrm{IC} \approx (B^2/8 \pi)/u_{rad}$, 
where $B^2 / 8 \pi$ is the magnetic field energy density 
and $u_{rad}$ is the radiation energy density responsible 
for the IC up-scattering. 
Assuming 1) that the X-ray flux of 5.5$\times 10^{-13}$~erg$\,$cm$^{-2}\,$s$^{-1}$ 
found by \citet{eger2010} originates from IC emission,
2) that the MEC applies, giving a magnetic 
field $B = 10.3$~$\mu$G,
3) that the radiation field is $u_{rad} = 40$~eV/cm$^3$ \citep{eger2010}, 
%then a SR flux $\nu F(\nu) \approx$~4$\times$10$^{-14}$~erg$\,$cm$^{-2}\,$s$^{-1}$ 
%is expected. This gives 
a flux density of the order of 300~kJy 
at the related synchrotron frequency of about 12~kHz is expected.
Extending a power-law spectrum up to the GHz range would 
require a spectral index around $-1.0$ for region~1.

The apparent validity of the IC emission hypothesis
for the diffuse X-ray emission found by \citet{eger2010}
could be further tested with an accurate radio spectral index.

\section{Conclusion}

In this paper the surroundings of the GC Terzan~5 
have been searched for indications of non-thermal diffuse emission. 
Several structures were isolated 
in the Effelsberg 21 and 11~cm surveys.
The currently available data point towards 
a complex radio emitting region. However, interpretation
of the radio structures is limited in the absence of reliable radio index. 
Future observations with high-frequency single-dish radio-telescopes
or a dedicated re-analysis of the original Effelsberg data would improve this situation.

The most intriguing of these structures extends from the position 
of the GC to about 0.8$^{\circ}$ to the north-west.
The energetics of this structure 
could be produced on a reasonable time scale 
by the population of msPSRs in Terzan~5 in a PWN scenario. 
Comparison of these time scales with the GC proper motion
would provide a strong test to this model.

A smaller compact radio feature may be associated
with an O~star. Indication for a molecular gas cavity
and order of magnitude agreement between the radio flux and
a Bremsstrahlung model support this scenario.
Confirmation of the molecular gas feature and of its distance 
are necessary tests to the proposed association.

There is an intriguing ridge-like extended radio source 
aligned with Terzan~5 roughly 1$^{\circ}$ north-east 
of the GC. In terms of energy content and spatial extent 
it shows similar characteristics to proposed 
remnants of highly asymmetric outbursts,
either from a micro-quasar or the final explosion 
of a star. Since a GC may be the host 
of highly asymmetric explosions, such an origin cannot be excluded.

The presented multi-wavelength data is compatible with an IC origin for 
the recently discovered extended diffuse X-ray emission 
from the direction of Terzan~5. A non-thermal Bremsstrahlung scenario
is not supported, although higher resolution data may change this conclusion.

Likewise, several negative results were obtained.
No significant density of molecular material in the environment 
of Terzan~5 could be found,
nor any clear hint of peculiar dust emission structure  
in the (low resolution) infrared data. The extension of  the
diffuse X-ray emission was limited 
to a 2\arcmin.5 radius region around Terzan~5,
to the sensitivity limit of the data. 

\begin{acknowledgements}

The authors thank Karl-Heinz Mack 
(INAF, Istituto di Radioastronomia Bologna, Italy)
for discussions about the radio analysis
and Matthieu Renaud (CNRS, LPTA Montpellier, France)
for sharing his expertise on \emph{INTEGRAL}.
This work was supported in part by the 
Polish Ministry of Science and Higher Education
project N N203 380336.

\end{acknowledgements}


\begin{thebibliography}{}

\bibitem[Abdo et al.(2009)]{abdo2009} Abdo, A.A. et al. (The Fermi LAT collaboration), 2009, Science, 325, 845A

\bibitem[Abdo et al.(2010)]{abdo2010} Abdo, A.A. et al. (The Fermi LAT collaboration), 2010, A\&A, 524, 75

\bibitem[Aharonian(2004)]{aharonian2004} Aharonian F.A. 2004, 'Very high energy cosmic gamma radiation : a crucial window on the extreme Universe', World Scientific Publishing

\bibitem[Baars et al.(1977)]{baars1977} Baars, J.W.M. et al. 1977, A\&A, 61, 99 

\bibitem[Bednarek \& Sitarek(2007)Bednarek,Sitarek]{bednarek2007} Bednarek, W. \& Sitarek, J. 2007, MNRAS, 377, 920

\bibitem[Beichman et al.(1988)]{beichman1988} Beichman C. et al. 1988,
'IRAS Catalogs and Atlases Explanatory Supplement', NASA RP-1190, vol 1

\bibitem[Blackburn(1995)]{blackburn1995} Blackburn J.K. 1995, ASP
Conf. Ser., vol. 77 

\bibitem[Boese(2000)]{boese2000} Boese, F.G. 2000, A\&AS, 141, 507B

\bibitem[Churazov et al.(2001)]{churazov2001} Churazov, E. et al. 2001, ApJ, 554, 261

\bibitem[Clemens(1985)]{clemens1985} Clemens, D.P. 1985, ApJ, 295, 422C

\bibitem[Condon et al.(1998)]{condon1998} Condon J.J. et al. 1998, AJ, 115, 1693
	
\bibitem[Dame et al.(2001)Dame,Hartmann,Thaddeus]{dame2001} Dame, T.M., Hartmann, D. \& Thaddeus, P., 2001, ApJ, 547, 792D

\bibitem[David et al.(1997)]{david1997} David L.P. et al., 1997, 
'The ROSAT High Resolution Imager calibration report',
Harvard-Smithsonian Center for Astrophysics,
ftp://legacy.gsfc.nasa.gov/rosat/doc/hri/hri\_report

\bibitem[Domainko \& Ruffert(2005)Domainko,Ruffert]{domainko2005} Domainko, W. \& Ruffert, M. 2005 A\&A, 444, L33

\bibitem[Domainko \& Ruffert(2008)Domainko,Ruffert]{domainko2008} Domainko, W. \& Ruffert, M. 2008, AdSpR, 41, 518

\bibitem[Draine \& Li(2007)]{draine2007} Draine, B.T. \& Li, A. 2007, ApJ, 657, 810D

\bibitem[Eger et al.(2010)Eger,Domainko,Clapson]{eger2010} Eger, P., Domainko, W. \&  Clapson, A.-C. 2010, A\&A, 513, 66

\bibitem[Everett et al.(2008)]{everett} Everett, J.E. et al. 2008, ApJ, 674, 258

\bibitem[Fender et al.(1999)]{fender1999} Fender, R.P., Garrington, S.T., McKay, D.J. et al. 1999, MNRAS, 304, 865

\bibitem[Ferraro et al.(2009)]{ferraro2009} Ferraro, F.R., Beccari, G., Dalessandro, E., et al. 2009, Nature, 462, 1028

\bibitem[Fruchter \& Goss(2000)]{fruchter2000} Fruchter A.S. \& Goss W.M. 2000, ApJ, 536, 865

\bibitem[Fukui et al.(2009)]{fukui2009} Fukui, Y., Furukawa, N., Dame, T.M. et al. 2009, PASJ, 61, L23 

\bibitem[Gaensler \& Slane(2006)]{gaensler2006} Gaensler, B.M. \& Slane, P.O. 2006, ARA\&A, 44, 17

%\bibitem[Garmire et al.(2003)]{garmire2003} Garmire, G.P. et al. 2003, 
%in Society of Photo-Optical Instrumentation Engineers (SPIE)
%Conf. Series, Vol. 4851, ed. J.E. Truemper \& H.D. Tananbaum, 28
\bibitem[Green(2009)]{green2009} Green, D.A. 2009, BASI, 37, 45

\bibitem[Grindlay et al.(2006)Grindlay,Portegies,McMillan]{grindlay2006} Grindlay, J., Portegies Z.S. \& McMillan, S. 2006, Nature Physics, 2, 116

\bibitem[Harris(1996)]{harris96} Harris, W. E. 1996, AJ, 112, 1487

\bibitem[Heinke et al.(2002)]{heinke2002} Heinke, C.O. et al. 2002, Bulletin of the American Astronomical Society, Vol. 34, p.1313

\bibitem[Heinke et al.(2006)]{heinke2006} Heinke, C.O. et al. 2006, ApJ, 651, 1098

\bibitem[Isobe et al.(2002)]{isobe2002} Isobe, N. et al. 2002, ApJ, 580, L111

\bibitem[Ivanova et al.(2008)]{ivanova2008} Ivanova, N., Heinke, C.O., Rasio, F.A., Belczynski, K. \& Fregeau, J. M. 2008, MNRAS, 386, 553

\bibitem[Johnston et al.(1995)]{johnston1995}Johnston H.M., Verbunt,
F. \& Hasinger, G. 1995, A\& A, 298, L21

\bibitem[Karovska et al.(2001)]{karovska2001} Karovska M. et al. 2001, ASP Conference Proceedings, Vol. 238, 435

\bibitem[Klein et al.(2003)]{klein2003} Klein, U., Mack, K.-H., Gregorini, L. \& Vigotti, M. 2003, A\&A, 406, 579

\bibitem[Kong et al.(2010)]{kong2010} Kong, A.K.H., Hui, C.Y. \& Cheng, K.S. 2010, ApJ, 712, 36

\bibitem[Koyama et al.(1995)]{koyama1995} Koyama, K., Petre, R., Gotthelf, E.V., et al. 1995, Nature, 378, 255

\bibitem[Gopal-Krishna \& Steppe(1980)]{gopal-krishna1980} Gopal-Krishna \& Steppe, H. 1980, A\&A, 88, 354

\bibitem[Krivonos et al.(2010)]{krivonos2010} Krivonos, R. et al. 2010, A\&A, 519, 107

\bibitem[Krockenberger \& Grindlay(1995)]{krockenberger1995} Krockenberger, M. \& Grindlay, J.E., 1995, ApJ, 451, 200K

\bibitem[Landecker et al.(2010)]{landecker2010} Landecker, T.L. et al. 2010, A\&A, 520, 80

\bibitem[Lanzoni et al.(2010)]{lanzoni2010} Lanzoni B. et al. 2010, ApJ, 717, 653

\bibitem[Lesch et al.(1992)]{lesch1992} Lesch, H. \& Reich, W. 1992, A\&A, 264, 493

\bibitem[Longair(1992)]{longair1992} Longair, M.S., 1992, 'High energy astrophysics. Vol.1: Particles, photons and their detection', Cambridge University Press

\bibitem[Miley(1980)]{miley1980} Miley, G. 1980, ARA\&A, 18, 165

\bibitem[Miville-Desch\^ene \& Lagache(2005)Miville-Deschêne,Lagache]{miville2005} Miville-Deschênes M.-A. \& Lagache G. 2005, ApJSS, 157, 302

\bibitem[Nakanishi \& Yoshiabi(2003)Nakanishi,Yoshiabi]{nakanishi2003} Nakanishi, H. \& Yoshiaki S. 2003, PASJ, 55, 191N

\bibitem[O'Dea et al.(2009)]{odea2009} O'Dea, C.P. et al. 2009, A\&A, 494, 471

\bibitem[Ochsenbein et al.(2000)]{ochsenbein2000} Ochsenbein F., Bauer P. \& Marcout J. 2000, A\&AS 143, 221

\bibitem[Okada et al.(2007)]{okada2007} Okada, Y.,  Kokubun, M.,  Yuasa, T. \& Makishima, K. 2007, PASJ, 59, 727

\bibitem[Paladini et al.(2003)]{paladini2003} Paladini, R., Burigana, C., Davies, R.D. et al., 2003, A\&A, 397, 213

\bibitem[Perner et al.(2000)]{perner2000} Perner, R., Raymons, J. \& Loeb, A., 2000, ApJ, 533, 658

\bibitem[Pooley \& Hut(2006)Pooley,Hut]{pooley2006} Pooley, D. \& Hut, P. 2006, ApJ, 646, L143

\bibitem[Ransom(2008)]{ransom2008} Ransom, S.M. 2008, in IAU Symposium, Vol. 246, IAU Symposium, ed. E. Vesperini, M. Giersz, \& A. Sills, 291–300

\bibitem[Reed(2005)]{reed2005} Reed B.C. 2005, AJ, 130,1652

%\bibitem[Reich et al.(1984)]{reich1984} Reich, W. et al. 1984, A\&AS, 58, 197R

\bibitem[Reich et al.(1988)]{reich1988} Reich, W. et al. 1988, A\&A, 191, 303

\bibitem[Reich et al.(1990a)]{reich1990a} Reich, W., Reich, P. \& F\"urst, E. 1990, A\&AS, 83, 539R

\bibitem[Reich et al.(1990b)]{reich1990b} Reich, W., F\"urst, E.,
  Reich, P. \& Reif, K. 1990, A\&AS, 85, 633

\bibitem[Rybicki \& Lightman(1979)]{rybicki1979} Rybicki, G.B. \& Lightman, A.P. 1979, 'Radiative processes in astrophysics', New York, Wiley-Interscience
	
\bibitem[Aavage \& de Boer(1981)]{savage1981} Savage, B.D. \& de Boer, K.S. 1981, ApJ, 243, 460S

\bibitem[Shara \& Hurley(2002)Shara,Hurley]{shara2002} Shara, M.M. \& Hurley, J.R. 2002, ApJ, 571, 716

\bibitem[Junkes et al.(1992)]{junkes1992} Junkes, N., Fuerst, E. \& Reich, W. 1992, A\&A, 261, 289

\bibitem[Vasquez et al. (2010)]{vasquez2010} Vasquez, J., Cappa, C.E., Pineault, S. \& Duronea, N.U. 2010, MNRAS, 405, 1976

\bibitem[Venter \& de Jager(2008)Venter,de Jager]{venter2008} Venter, C. \& de Jager, O.C. 2008, AIPC, 1085, 277

\bibitem[Venter et al.(2009)Venter,de Jager,Clapson]{venter2009} Venter, C., de Jager, O.C. \& Clapson, A.-C. 2009, ApJ, 696, L52

\bibitem[Yamamoto et al.(2008)]{yamamoto2008} Yamamoto, H., Ito, S., Ishigami, S., et al. 2008, PASJ, 60, 715

\end{thebibliography}
\end{document}